\documentclass[12pt]{iopart}

\usepackage{iopams}
\usepackage{epsfig,graphics}
\begin{document}

\title[Noise influence on electron dynamics in semiconductors]
{Noise influence on electron dynamics in semiconductors driven by a periodic electric field}

\author{D. Persano Adorno$^a$, N. Pizzolato$^{ab}$\footnote{E-mail: npizzolato@gip.dft.unipa.it}
 and B. Spagnolo$^{ab}$\footnote{E-mail: spagnolo@unipa.it} }

\address{$^a$Dipartimento di Fisica e Tecnologie Relative and CNISM,\\
$^b$Group of Interdisciplinary Physics\footnote{URL: http://gip.dft.unipa.it}\\
 Viale delle Scienze, Ed. 18,Palermo, Italy}
\ead{dpersano@unipa.it}

\begin{abstract}
Studies about the constructive aspects of noise and fluctuations in
different non-linear systems have shown that the addition of external
noise to systems with an intrinsic noise may result in a less noisy
response. Recently, the possibility to reduce the diffusion
noise in semiconductor bulk materials by adding a random fluctuating
contribution to the driving static electric field has been tested.
The present work extends the previous theories by considering
the noise-induced effects on the electron transport
dynamics in low-doped n-type GaAs samples driven by a
high-frequency periodic electric field (cyclostationary conditions).
By means of Monte Carlo simulations, we calculate the changes in the spectral density of the electron
velocity fluctuations caused by the addition of an external
correlated noise source. The results reported in this paper confirm
that, under specific conditions, the presence of a
fluctuating component added to an oscillating electric field
can reduce the total noise power. Furthermore, we
find a nonlinear behaviour of the spectral density with the noise intensity.
Our study reveals that, critically depending on the external noise correlation time,
the dynamical response of electrons driven by a periodic electric field
receives a benefit by the constructive interplay between the fluctuating field
and the intrinsic noise of the system.

\end{abstract}

\maketitle

\section{Introduction}\label{sect1}
The presence of noise in experiments is generally considered a
disturbance, especially studying the efficiency of
semiconductor-based devices, where strong fluctuations could affect
their performance. Recently, however, an increasing interest has
been directed towards the constructive aspects of noise on the
dynamical response of non-linear systems. A counterintuitive
enhancement of the stability can be induced in systems containing
metastable states by the addition of noise \cite{Mante1996,
Agud2001, Fiasco2003, Fiasco2005}. Some excitable systems may
achieve more order in the presence of noise, even in the absence of
an external signal, such as in phenomena of coherent resonance where
an optimal level of noise leads to regular excursions from the
ground state \cite{Pikovsky}. The effect of interaction between an
external source of fluctuations and an intrinsically noisy system
has been analytically investigated, for the first time, by Vilar and
Rub\'{i} (2001). They have demonstrated that the spectral intensity
of the output signal in a low frequency domain can be reduced by the
addition of small amplitude noise on the input of the system
\cite{vilar}. Their analytical theory was developed by using very
general statistical properties of nonlinear systems and by assuming
internal and external sources of noise completely uncorrelated. The
main limitation of this theory is the assumption of an intrinsic
white noise. These restrictions persist in the recent work of Walton
and Visscher (2004), in which they extended the previous theory by
considering the total power spectrum of the output signal for
greater intensities of the input fluctuations \cite{walton}. In
realistic cases, however, there always exists some correlation
during the characteristic relaxation time of the system and the
noise spectra are not strictly white \cite{seol}.

In semiconductor bulk materials, the possibility to reduce the
diffusion noise by adding a correlated random contribution to a
driving static electric field, has been tested by Varani and
collaborators in 2005 \cite{varani}. Their numerical results,
obtained by including energetic considerations in the theoretical
analysis, have shown that, under specific conditions of the fluctuating electric
field, it is possible to suppress the intrinsic noise in n-type GaAs
bulk \cite{varani}.

Recent studies of the electron velocity fluctuations in GaAs bulks
driven by periodic electric fields, have shown that the spectral
density strongly depends on the frequency of the applied field and
critical modifications are observed when two mixed high-frequency
large-amplitude periodic electric fields are used \cite{persano081}.
This means that the total power spectrum of the intrinsic noise is
dependent on both the amplitude and the frequency of the excitation
signals \cite{persano081}. In the wake of these results, we have
found that the opportunity to suppress the diffusion noise in
semiconductor bulk materials exists also under cyclostationary
conditions, by adding a gaussian fluctuating contribution to the
driving electric field \cite{persano082}.

In this paper we investigate the noise-induced effects on the
intrinsic carrier noise spectral density in low-doped n-type GaAs
semiconductor driven by a high-frequency periodic electric field.
The electron dynamics is simulated by a Monte Carlo procedure which
keeps into account all the possible scattering phenomena of the hot
electrons in the medium. The semiconductor intrinsic noise is
obtained both by computing the velocity fluctuations correlation
function and the spectral density \cite{Shiktorov, Gonzalez} and
directly calculating the variance of the electron velocity
fluctuations. The effects caused by the addition of an external
source of correlated noise are investigated by analyzing (i) the
noise spectral density at the same frequency of the external driving
field and (ii) the integrated spectral density (ISD), which
coincides with the variance of the electron velocity fluctuations.
Our results confirm that the presence of
a random contribution to a high-frequency periodic electric field
can reduce the total noise power. Furthermore, we find a nonlinear
behaviour of both the noise spectral density at the driving
frequency and the ISD, with the noise intensity. In particular, the
ISD shows a minimum which critically depends on the value of the
noise correlation time. Detailed investigations of the electron
transport dynamics in the semiconductor reveal that the system
receives a benefit by the constructive interplay between the random
fluctuating electric field and the intrinsic noise. The paper is
organized as follows: in section \ref{sect2} we present the details
of the Monte Carlo procedure, the statistical quantities used to
investigate the electronic noise and a physical model describing the
dependence of the noise spectra on the frequency of the electric
field; in section \ref{sect3} the results of our calculations are
given and discussed. Final comments and conclusions are given in
section \ref{sect4}.

\section{Semiconductor model and noise calculation methods}\label{sect2}
The electron dynamics in a GaAs bulk semiconductor, driven by an
oscillating electric field, is simulated by a Monte Carlo method.
The motion of electrons is characterized by an average velocity,
which depends on the external parameters of the system, such as the
amplitude of the applied electric field and its frequency. The
fluctuations of electron velocity around its mean value correspond
to the intrinsic noise of the system.

\subsection{Monte Carlo procedure}\label{sect21}
The Monte Carlo algorithm, developed for simulating the motion of
electrons in a GaAs semiconductor, follows the standard procedure
described in Ref.~\cite{Persano00}. The conduction bands of GaAs are
the $\Gamma$ valley, four equivalent $L$-valleys and three
equivalent $X$-valleys. The parameters of the band structure and
scattering mechanisms are also taken from Ref.~\cite{Persano00}. Our
computations include the effects of the intravalley and intervalley
scattering of the electrons, in multiple energy valleys, and of the
nonparabolicity of the band structure. Electron scatterings due to
ionized impurities, acoustic and polar optical phonons in each
valley as well as all intervalley transitions between the equivalent
and non-equivalent valleys are accounted for. We assume
field-independent scattering probabilities; accordingly, the
influence of the external fields is only indirect through the
field-modified electron velocities. All simulations are obtained in
a GaAs bulk with a free electrons concentration $n=10^{13}$ cm$^3$.
To neglect the thermal noise contribution and to highlight the
partition noise effects we have chosen a lattice temperature $T=$ 80
K. We have assumed that all donors are ionized and that the free
electron concentration is equal to the doping concentration.

The semiconductor bulk is driven by a fluctuating periodic electric
field
\begin{eqnarray}
E(t)=E_{\rm 0}\cos({\omega t+\phi})+\eta(t) \label{eq1}
\end{eqnarray}
with frequency $f=\omega / 2 \pi$ and amplitude $E_{\rm 0}$. The
random component of the electric field is modeled with an
Ornstein-Uhlenbeck (OU) stochastic process $\eta(t)$, which obeys
the following stochastic differential equation:
\begin{eqnarray}
\frac{d\eta(t)}{dt}=-\frac{\eta(t)}{\tau_{\rm c}}+\sqrt{\frac{2D}{\tau_{\rm c}}}\xi(t)
\label{eq2}
\end{eqnarray}
where $\tau_{\rm c}$ and $D$ are, respectively, the correlation time
and the variance of the OU process, and $\xi(t)$ is the Gaussian
white noise with the autocorrelation
$<\xi(t)\xi(t^\prime)>=\delta(t-t^\prime)$. The OU correlation
function is $<\eta(t)\eta(t^\prime)>=D \exp{(-|t-t^\prime|/\tau_{\rm
c})}$.

\subsection{Semiconductor noise calculation}\label{sect22}
The changes on intrinsic noise properties are investigated by the
statistical analysis of the autocorrelation function of the velocity
fluctuations and of its mean spectral density. When the system is
driven by a periodic electric field (cyclostationary conditions),
the correlation function $C_{\delta v\delta v}(t,\tau)$ of the
velocity fluctuations $\delta v(t)=v(t)- \left\langle v
(t)\right\rangle$ can be calculated \cite{Gonzalez} as
\begin{equation}
  C_{\delta v\delta v}(t,\tau) =\left\langle
  v\left(t-\frac{\tau}{2}\right)v\left(t+\frac{\tau}{2}\right)\right\rangle-\left\langle
  v\left(t-\frac{\tau}{2}\right)\right\rangle\left\langle v\left(t+\frac{\tau}{2}\right)\right\rangle
\label{eq4}
\end{equation}
in which $\tau$ is the correlation time and the average is meant
over a sequence of equivalent time moments $t=s+mT$, with
$s$ belonging to the time interval $[0, T]$ ($T$ is
the field period) and $m$ is an integer \cite{Gonzalez}. This
two-time symmetric correlation function eliminates any regular
contribution and describes only the fluctuating part of $v(t)$. By
averaging over the whole set of values of $t$ within the period
$T$, the velocity autocorrelation function becomes
\begin{equation}
C_{\delta v\delta v}(\tau) =\frac{1}{T}\int_{0}^{T}C_{\delta
v\delta  v}(t,\tau)dt \label{eq5}
\end{equation}
and the spectral density can be calculated as the Fourier transform
of $C_{\delta v\delta v}(\tau)$. In the computations of the
autocorrelation function we have considered $10^3$ possible initial
values of $s$ and a total number of equivalent time moments $m\cong
10^6$.

Intrinsic noise has been investigated also by estimating directly
the electron velocity variance. This calculation has been performed
separately for each energy valley, following the same method of
equivalent time moments described above.

\subsection{Physical model of intrinsic noise}\label{sect23}
When the semiconductor is driven by a static electric field, the
shape of the spectral density of electron velocity fluctuations is
exclusively determined by the strength of the applied field. For
amplitudes smaller than the threshold field (Gunn Field) $E_{\rm G}$
for intervalley transitions, the diffusion is the most relevant
source of noise, while, for $E>E_{\rm G}$, the complex structure of
the semiconductor becomes relevant and random transitions of
carriers among the available energy valleys must be taken into
consideration. In this case, the intrinsic noise is mainly
determined by a partition noise, caused by stochastic carrier
transitions between regions characterized by different dynamical
properties (intervalley transfers) in momentum space. The partition
noise is characterized by a pronounced peak in the spectral density
at a frequency $\nu_{\rm G}$, which is defined as the "natural"
transition frequency of the system between the valleys
\cite{persano081}.

Under cyclostationary conditions, the noise behaviour depends on
both the amplitude and the frequency of the applied field. In
particular, it is similar to that of the static field case only for
very low-frequency fields ($f\ll\nu_{\rm G}$). On the contrary, for
frequencies $f \gtrsim \nu_{\rm G}$, the intervalley transfers are
driven by the external field, the system enters in a forced regime
of oscillations and the velocity fluctuations become time correlated
\cite{persano081}. In this case, the spectral density exhibits: (i)
a peak centered around the frequency of the periodic signal and (ii)
a significant enhancement in the low-frequency region.

In this work we focus our attention to the noise induced effects on
the electron dynamics in a GaAs semiconductor, driven by an
high-frequency oscillating field.

\section{Numerical results and discussion}\label{sect3}
\begin{figure}[htbp]
\includegraphics [width=16cm,height=8cm]{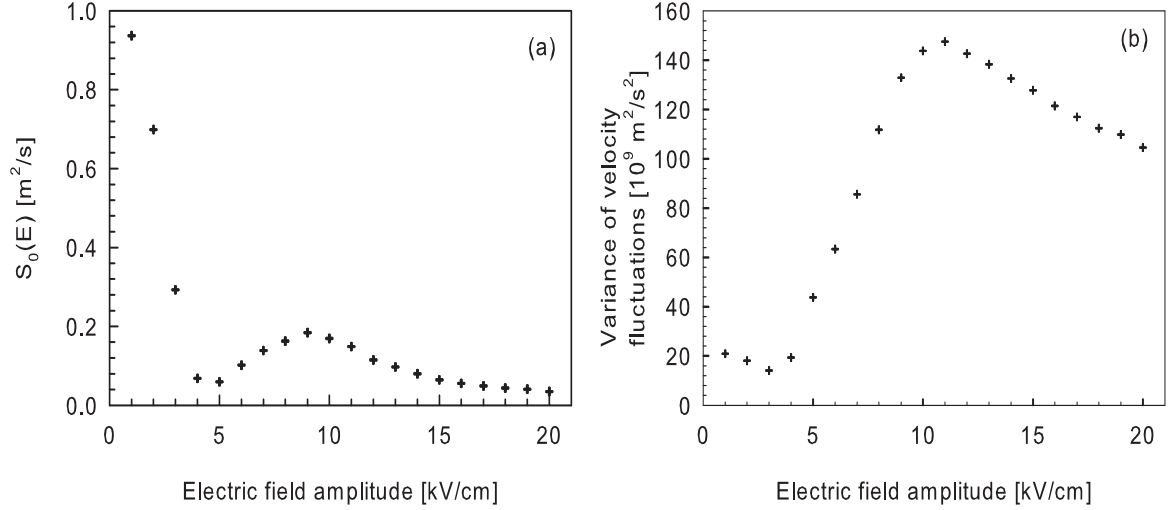}
\caption{(a) Spectral density $S_0(E)$ at zero frequency and (b)
variance of electron velocity fluctuations as a function of the
amplitude of the oscillating electric field.} \label{fig1}
\end{figure}

The spectral density of the electron velocity fluctuations has been
studied by adopting a fluctuating periodic electric field with
frequency $f=500$ GHz. The amplitude of this field has been chosen
on the base of a preliminary analysis on the variance of velocity
fluctuations and the spectral density $S_0(E)$ at zero frequency, as
a function of the amplitude of the oscillating field. Following
Ref.~\cite{vilar} and Ref.~\cite{varani}, the most favorable
condition to obtain a noise suppression effect in our system is
reached when $d^2S_0(E)/dE^2$ is negative and the variance of
velocity fluctuations exhibits a maximum. In figure \ref{fig1}a and
\ref{fig1}b we can see which range of amplitudes of electric field
verifies these conditions. Accordingly, we have chosen a driving
electric field with amplitude $E_0=10$ kV/cm and frequency $f=500$
GHz.

\begin{figure}[htbp]
\includegraphics [width=16cm,height=8cm]{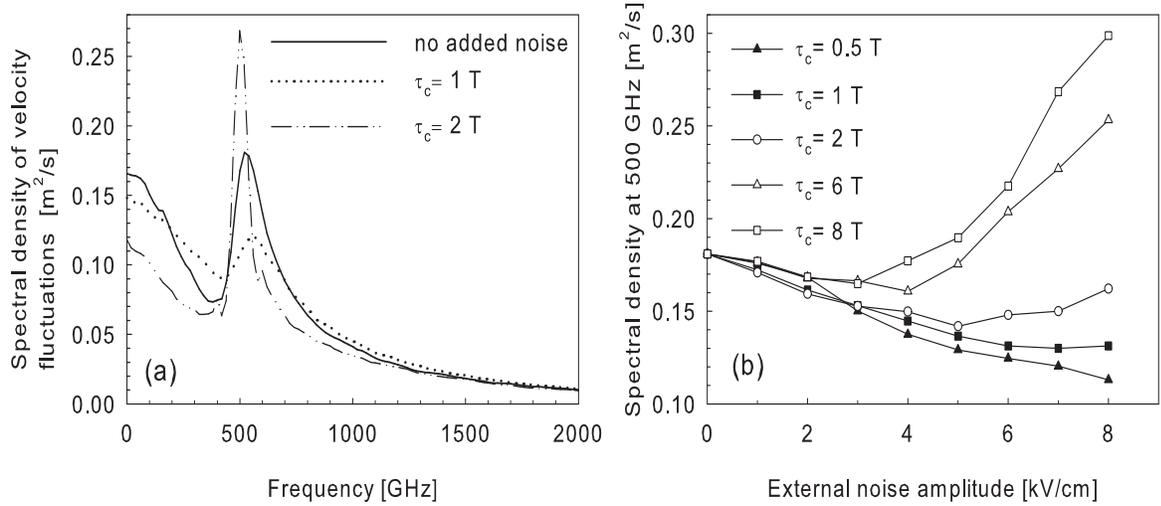}
\caption{(a) Spectral density of electron velocity fluctuations as a
function of the frequency. Solid line is obtained in the absence of
external noise; dotted line describes the results obtained with
$D^{1/2}=7$ kV/cm and $\tau_{\rm c}=1$ ps $=0.5$ $T$; dashed-dotted
line is obtained with $D^{1/2}=7$ kV/cm  and $\tau_{\rm c}=16$ ps $=8$ $
T$. (b) Height of the peak in the spectral density of electron
velocity fluctuations as a function of the external
noise amplitude for five different values of the correlation time $\tau_{\rm
c}$ of the added noise source.} \label{fig2}
\end{figure}

In the absence of external noise, the amplitude of the forcing field
is large enough to switch on intervalley transitions from the
$\Gamma$ valley to the $L$ valleys and, since the frequency $f$ is
of the same order than $\nu_{\rm G}$, the electron velocity
fluctuations are manly determined by partition noise
\cite{Shiktorov03}. In this case, the spectrum is characterized by
the features described in section \ref{sect23}. In figure
\ref{fig2}a we show how the spectral density of electron velocity
fluctuations is modified by the presence of noise. The addition of
an external source of fluctuations to the driving electric field
strongly changes the spectrum and, in particular, the height of the
peak around $500$ GHz, in a way that critically depends on the OU
correlation time.
In figure \ref{fig2}b we plot the maximum of the spectral density at
the frequency of the driving field as a function of the external
noise amplitude $D^{1/2}$, for five different values of $\tau_{\rm
c}$. An interesting nonlinear behavior of this quantity is observed
for increasing noise intensities and correlation times. In
particular, for values of $\tau_{\rm c}$ smaller than or equal to the
period $T$ of the oscillating electric field, the spectral density
at $500$ GHz shows a monotonic decreasing trend with increasing
noise amplitude. For values of $\tau_{\rm c}$ greater than $T$, the
spectral density is reduced only for small amplitudes of the
external noise, while an enhancement of the peak is observed for
greater intensities. When the intrinsic noise is mainly due to the
partition effect, the height of the peak in the spectral density
depends on the population of the different valleys, reaching a
maximum when the populations are nearly at the same level
\cite{Shiktorov03,nougier}. Since the "effective" electric field
experienced by electrons in the presence of a fluctuating field is
different, the number of intervalley transitions changes with
respect to the case in which the external source of noise is absent.
This fact can be responsible of the observed changes on the peak of
the spectral density.

\begin{figure}[htbp]
\includegraphics [width=13cm,height=9cm]{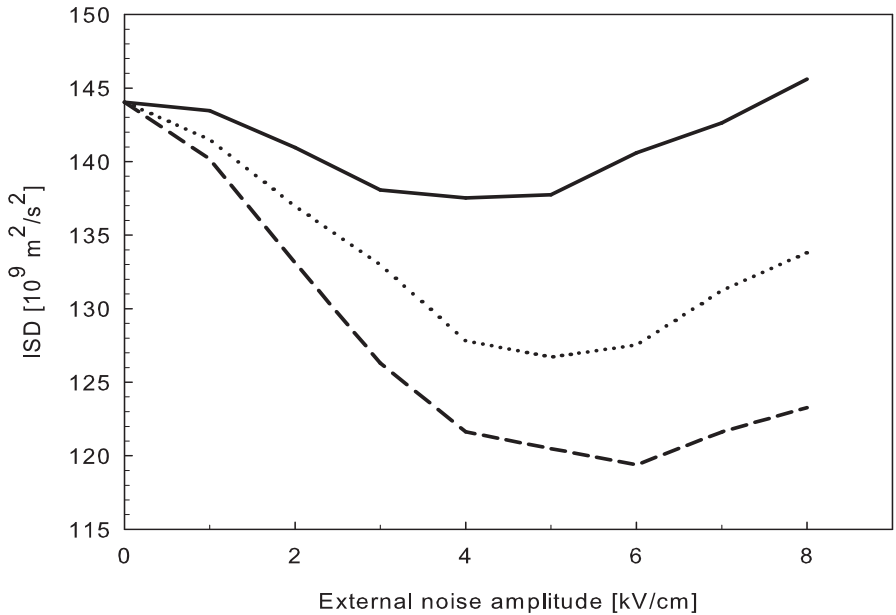}
\caption{Integrated spectral density of electron velocity
fluctuations as a function of the external noise amplitude. Solid
line: $\tau_{\rm c}=0.5$ $T$; dotted line: $\tau_{\rm c}=2$ $T$; dashed line: $\tau_{\rm c}=8$ $T$.} \label{fig3}
\end{figure}

The dependence of the intrinsic noise suppression effect on the
amplitude and the correlation time of the external source of
fluctuations has been investigated also by studying the integrated
spectral density (ISD), i.~e.~the total noise power, as a function
of the OU noise amplitude, for three different values of $\tau_{\rm
c}$, namely 0.5, 2 and 8 $T$. In figure \ref{fig3} we show a clear
reduction of the ISD in the presence of external noise. In
particular, for each value of the correlation time we find a range
of $D^{1/2}$ in which the electric field fluctuations reduce the
semiconductor intrinsic noise. This effect is more evident for
higher correlation times.

\begin{figure}[htbp]
\includegraphics [width=11cm,height=20cm]{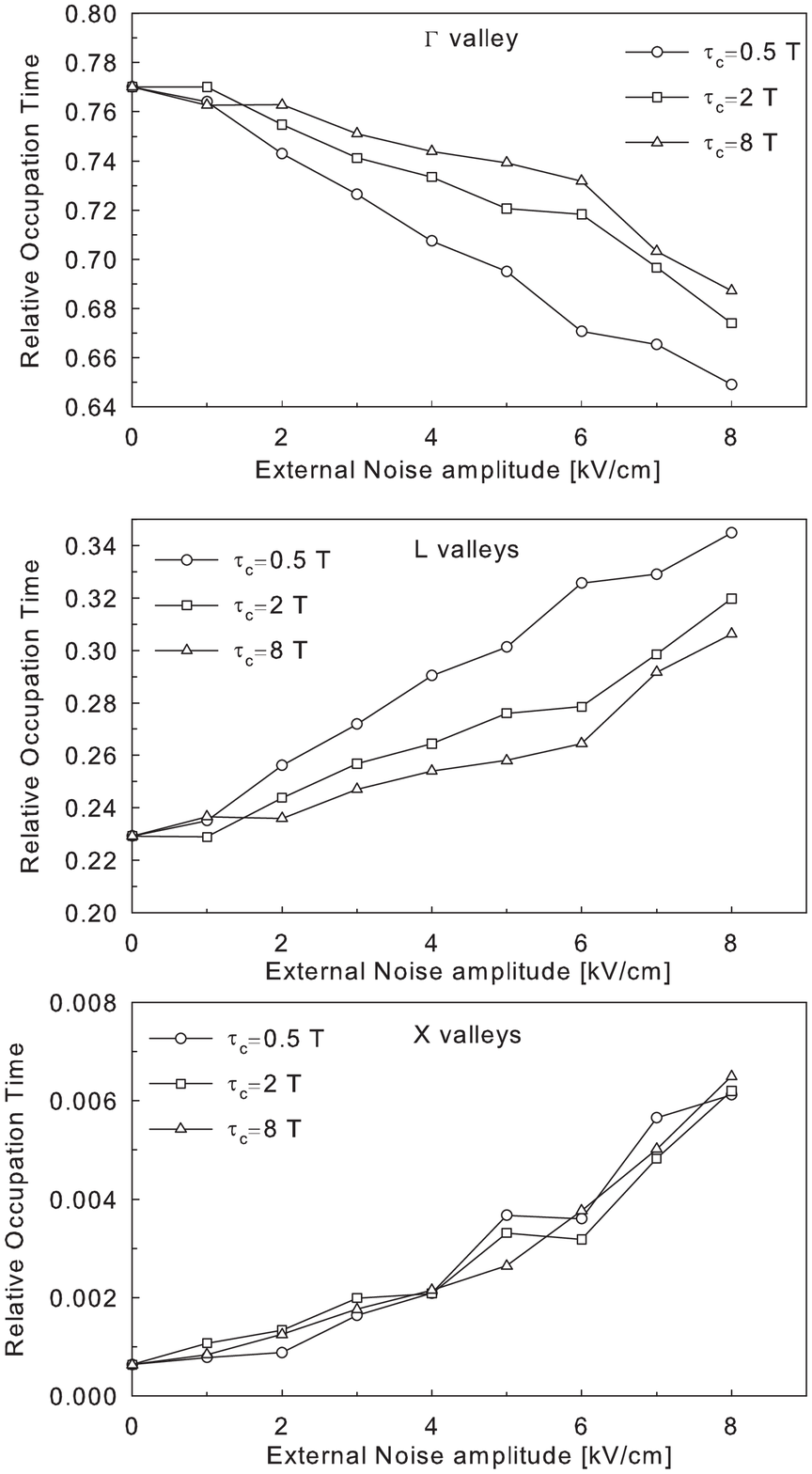}
\hspace{-3.0cm}
\includegraphics [width=11cm,height=20cm]{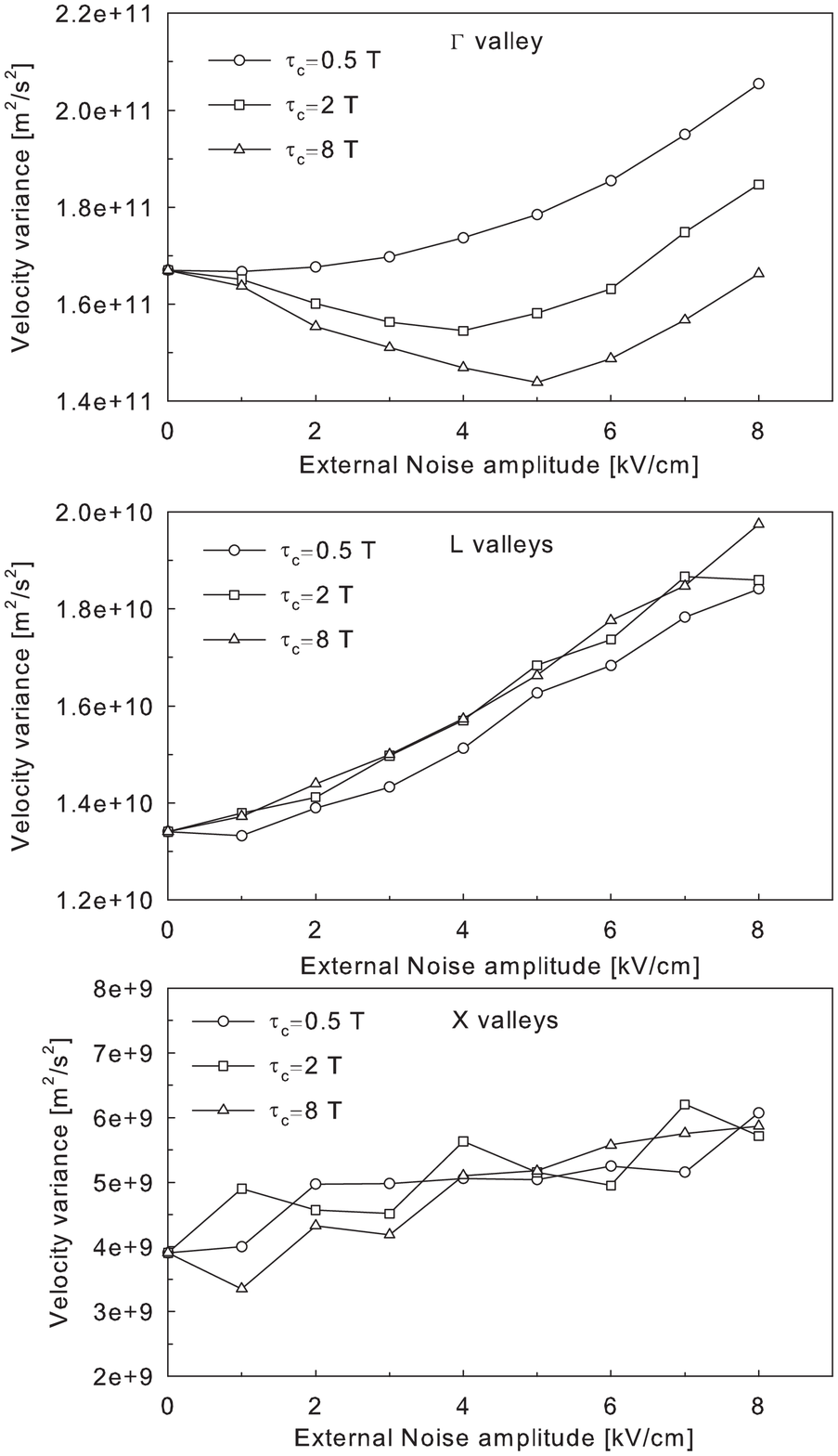}
\caption{Relative occupation time (left panels) and  variance of the electron velocity fluctuations
(right panels) as a function of the
external noise amplitude, for three different correlation times.} \label{fig4}
\end{figure}

From a microscopic point of view, this suppression can arise
from the fact that the fluctuating electric field forces the
carriers to visit regions of the momentum space characterized by a
smaller variance with respect to the case of zero noise \cite{walton}. We have
investigated the details of the electron dynamics under the
fluctuating electric field by analyzing the relative occupation time
and the velocity variance separately in different valleys, for
different correlation times. In figure \ref{fig4} (left panels) we
show that, when the noise intensity increases, the electron
occupation time of the $\Gamma$ valley decreases and the
corresponding times calculated for the $L$ and $X$ valleys increase.
This behaviour is expected because the addition of fluctuations to
the driving electric field leads to an increase of scattering events
which are responsible for an increase of transitions from the
$\Gamma$ valley to valleys at higher energy. Moreover, this
behaviour depends on the correlation time of the external noise
source. In particular, for any fixed value of the external noise
amplitude, the effect of reduction of the relative occupation time
for the $\Gamma$ valley and the corresponding increase for the $L$
valleys is more pronounced for shorter correlation times.

Less obvious is the behaviour of the electron velocity variance
evidenced in figure \ref{fig4} (right panels). In fact, while the
common experience would suggest an increase of the velocity variance
when the external noise intensity grows up, we find that the
velocity variance in the $\Gamma$ valley can be reduced in a
specific range of the noise amplitude, depending on the value of the
correlation time. An increasing trend is instead observed for the
$L$ and $X$ valleys. The reduction of the electron velocity variance
observed in the $\Gamma$ valley for $\tau_{\rm c}= 2$ $T$ and $D^{1/2}$
between 1 and 6 and, even more, for $\tau_{\rm c}= 8$ $T$ and $D^{1/2}$
between 1 and 8, represents an intrinsic effect of the dynamics of
electrons in the $\Gamma$ valley without taking into account any
transfer to valleys characterized by different dynamical properties.
This effect of noise-induced stability can explain the longer
residence times of electrons in the $\Gamma$-valley at higher
correlation times.

\section{Conclusions}\label{sect4}
The results reported in this work confirm that the intrinsic noise
in a n-type GaAs semiconductor can be reduced by the addition of
external fluctuations to the driving high-frequency periodic
electric field. These findings have been obtained by investigating
the noise-induced modifications of the spectrum of electron velocity
fluctuations and the ISD. The reduction of the characteristic peak
(at the driving frequency) of the noise spectral density has been
observed for a wide range of noise amplitudes and for OU correlation
times shorter than the period of the driving electric field. For
higher values of $\tau_{\rm c}$, a nonmonotonic behaviour of this
maximum is clearly evident. A less noisy response in the presence of a driving
periodic electric field containing time-correlated fluctuations is observed.
This interpretation is confirmed by our study on the electron
velocity variance, calculated separately for every single energy
valley of the semiconductor. Previous studies ascribe the reduction of
the electron velocity fluctuations to an overall effect of
intervalley transfers. In this work, we have shown that the velocity
variance of an electron moving in the $\Gamma$-valley is reduced by
the presence of correlated noise, independently from the transitions
to upper valleys, bringing to longer residence times. This effect of
noise enhanced stability (NES) arises from the fact that the
transport dynamics of electrons in the semiconductor receives a
benefit by the constructive interplay between the fluctuating
electric field and the intrinsic noise of the system.

To conclude, both the amplitude and the correlation time of the
electric field fluctuations are important for the intrinsic noise
reduction effect. Further studies are needed to investigate the existence of a
relationship between the semiconductor characteristic time scales,
which essentially depend on the scattering probabilities, the
noise correlation time and the period of the oscillating driving field.

\ack
This work was partially supported by MIUR and CNISM-INFM.

\end{document}